Towards a Biomechanical Understanding of Tempo in the Golf Swing

Robert D. Grober
Department of Applied Physics
Yale University

Jacek Cholewicki
Department of Biomedical Engineering
Yale University

It is proposed that aspects of the tempo of the golf swing can be understood in terms of a biomechanical clock. This model explains several aspects of tempo in the golf swing; including total duration of the golf swing, the ratio of backswing to downswing time, and the relative insensitivity of tempo on the length of the golf shot. We demonstrate that this clock and the resulting tempo are defined by of the rotational inertia of the body/club system and the elastic properties of the body, yielding a system which can be modeled as a simple harmonic oscillator.

Tempo refers to the pace of the golf swing. It is can be characterized by measuring the duration of the backswing, $T_b$, and the duration of the downswing, $T_d$. While qualitative discussions of tempo are as old as the game itself, quantitative measurement of tempo has existed only for a couple of decades [1]. Recently, a study of the tempo of professional golfers was published in the book *Tour Tempo* [2] in which it was pointed out that the ratio of backswing to downswing time of professional golfers is of order three, $T_b/T_d \approx 3$. These measurements were made using the frame rate of standard video (i.e. 30 Hz frame rate) as the clock. The tempo of the majority of tour professionals studied in *Tour Tempo* is characterized by $T_b \approx 24$ frames and $T_d \approx 8$ frames. The ratio $T_b/T_d$ for all players reported in the study covered the range from 21/7 to 30/10. Also pointed out in *Tour Tempo* is that the overall tempo of professional golfers is significantly faster than that of the average golfer and that the tempo does not change significantly with the length of the shot or the type of club.

In an attempt to more thoroughly characterize tempo, we have performed measurements on a wide variety of golfers, from tour professionals to the average weekend golfers, using electronics embedded in the shaft of the golf club. The details of this measurement system are described elsewhere [3]. In summary, motion sensing accelerometers and wireless communications electronics mounted in the shaft allow us to determine the start of the swing, the transition from backswing to downswing, and the point of impact. Sampling rates are of order 250 Hz, yielding eight times more detail than that obtained from conventional video.

Ten to twenty swings were recorded for each golfer as they hit a five iron. $T_b$ and $T_d$ are measured for each swing. Using the ensemble of swings, the average and standard deviation of $T_b$ and $T_d$ is calculated for each golfer. The standard deviation is used to characterize reproducibility. The results of these measurements are displayed in Figs. 1. The golfers are divided into three groups: a) playing professionals (*n*=12); b) teaching professionals and good amateur golfers (*n*=13); and c) all other golfers (*n*=18). Each data point represents the averaged result for a particular golfer. The standard deviations are indicated by error bars.

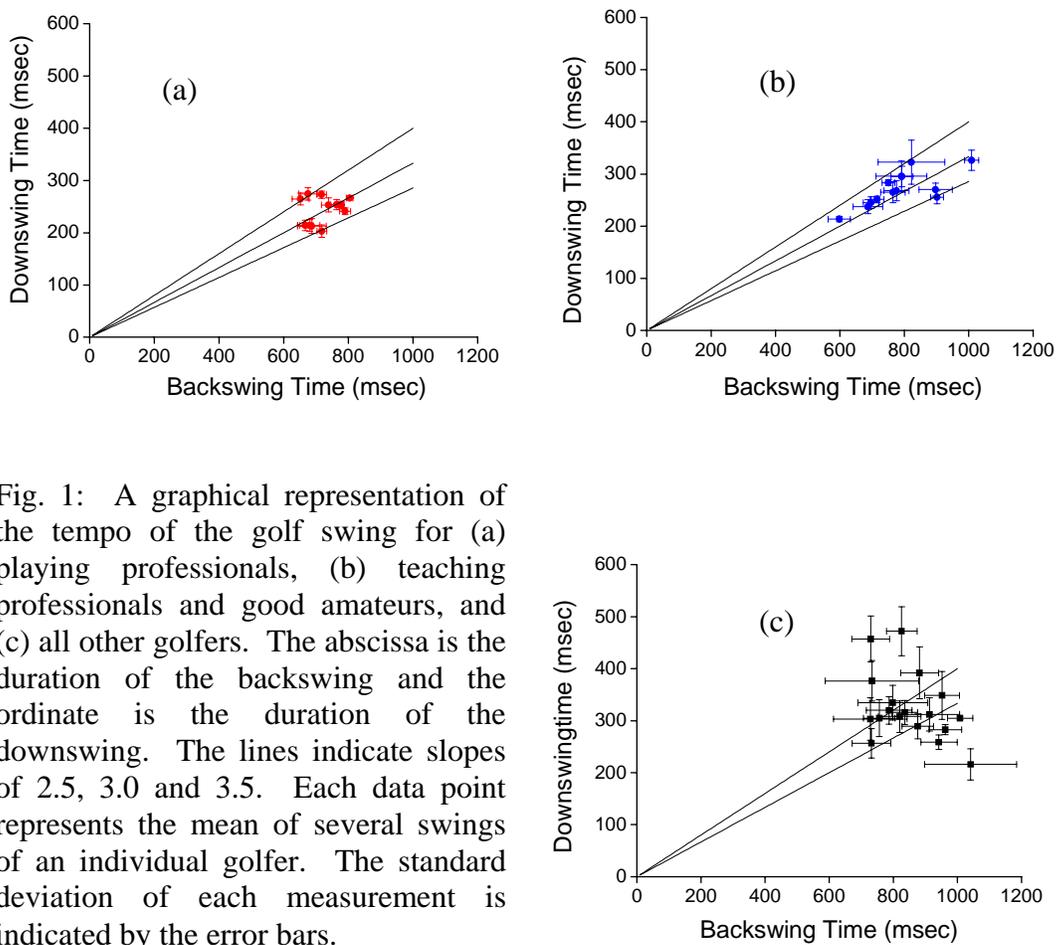

Fig. 1: A graphical representation of the tempo of the golf swing for (a) playing professionals, (b) teaching professionals and good amateurs, and (c) all other golfers. The abscissa is the duration of the backswing and the ordinate is the duration of the downswing. The lines indicate slopes of 2.5, 3.0 and 3.5. Each data point represents the mean of several swings of an individual golfer. The standard deviation of each measurement is indicated by the error bars.

The data for playing professionals is consistent with the data reported in *Tour Tempo*. The ratio $T_b/T_d$ varies between 2.5 and 3.5, with the average being nearer to 3.0. Additionally, these golfers exhibit very small values of standard deviation relative to all other golfers, meaning that their swings are very reproducible. Note that the time of the backswing of the tour professionals all seem to cluster in the vicinity of $T_b \approx 0.7 - 0.8$ sec, which corresponds to 21-24 frames per second of standard video, again consistent with the data presented in *Tour Tempo*.

The distributions in Fig 1(b) and Fig 1(c) are noticeably larger than that of Fig 1(a). In both cases, the average ratio $T_b/T_d$ is centered at 3.0, but varies widely, much more so for the last group of golfers. As one might expect, the standard deviations measured for most other golfers are much larger than that of tour professionals. Additionally, the golf swings of the professional golfer are uniformly faster than that of the average golfer.

The following three aspects of the tempo of playing professionals suggest the workings of a biomechanical clock: 1) the fact that all playing professionals have roughly the same tempo, 2) the very small variance in tempo from swing to swing for individual playing professionals, and 3) the relative insensitivity of tempo to the length of the swing or the type of club. In this section we explore the plausibility that the simplest model of a clock, the simple harmonic oscillator, can be used to understand tempo in the golf swing of professional golfers.

The simple harmonic oscillator requires a mass and a spring, *i.e.* a restoring force. In this proposed model the mass is comprised of the torso, legs, arms, and club. The spring results from the "effective" elasticity of the biomechanical system, comprising the natural and trained response of the body. The importance of elasticity in animal movement has long been documented [*4, 5, 6, 7, 8*], and we propose it plays a central role in defining the tempo of the golf swing of professional golfers.

As a test of plausibility, consider the equation of motion of the driven harmonic oscillator, $m\ddot{x}(t) + kx(t) = F(t)$, where $m$ is the mass, $k$ is the spring constant, $x$ is the displacement and $F$ is the driving force. The initial conditions of the backswing are $x_b(0) = 0$ and $\dot{x}_b(0) = 0$. The backswing begins with the application of a constant force $F_B$, causing the spring to compress subject to the equation $\ddot{x}_b(t) + \Omega^2 x_b(t) = \frac{F_b}{m}$, where $\Omega = \sqrt{\frac{k}{m}}$ is the resonant frequency of the oscillator. The solution is given as

$$x_b(t) = \frac{F_b}{k}(1 - \cos\Omega t)$$

The spring is maximally compressed after a period of time, $T_b = \frac{\pi}{\Omega}$, independent of the applied force. At this moment of maximum compression the position of the mass is $\frac{2F_b}{k}$ and the velocity is zero. Thus, the duration of the backswing is independent of the applied force, while the maximum displacement of the backswing is proportional to the applied force.

The downswing begins at this point of maximum compression. The direction of the applied force is reversed so as to help decompress the spring and return the mass to the original position. We now solve this new problem $\ddot{x}_d(t) + \Omega^2 x_d(t) = -\frac{F_d}{m}$ subject to the initial conditions $x_d(T_b) = \frac{2F_b}{k}$ and $\dot{x}_d(T_b) = 0$. The solution to this problem is given as

$$x_d(t) = -\frac{F_d}{k} + \frac{2F_b + F_d}{k}\cos\Omega(t - T_b)$$

One can solve this equation for the time at which the club returns to the origin, $x_d(T_b + T_d) = 0$, yielding the transcendental equation

$$\cos \Omega(T_d) = \frac{F_d}{F_d + 2F_b}$$

One can now solve for the ratio $T_b/T_d$ as a function of $F_d/F_b$

$$\frac{T_b}{T_d} = \frac{\pi}{\cos^{-1}\left(\frac{F_d/F_b}{F_d/F_b + 2}\right)}$$

This equation is plotted in Fig. 2, where the ratio $T_b/T_d$ is shown as a function of $F_d/F_b$. The points $F_d/F_b = 1.0, 2.0, 3.0,$ and $4.0$ are indicated as asterisks. It is interesting to note that the ratio $T_b/T_d$ falls in the range $2.5 < T_b/T_d < 3.5$ as long as the ratio $F_d/F_b$ is constrained to $1 < F_d/F_b < 3$. As it relates to the results presented in *Tour Tempo*, this model yields $T_b/T_d = 3$ when $F_d/F_b = 2$.

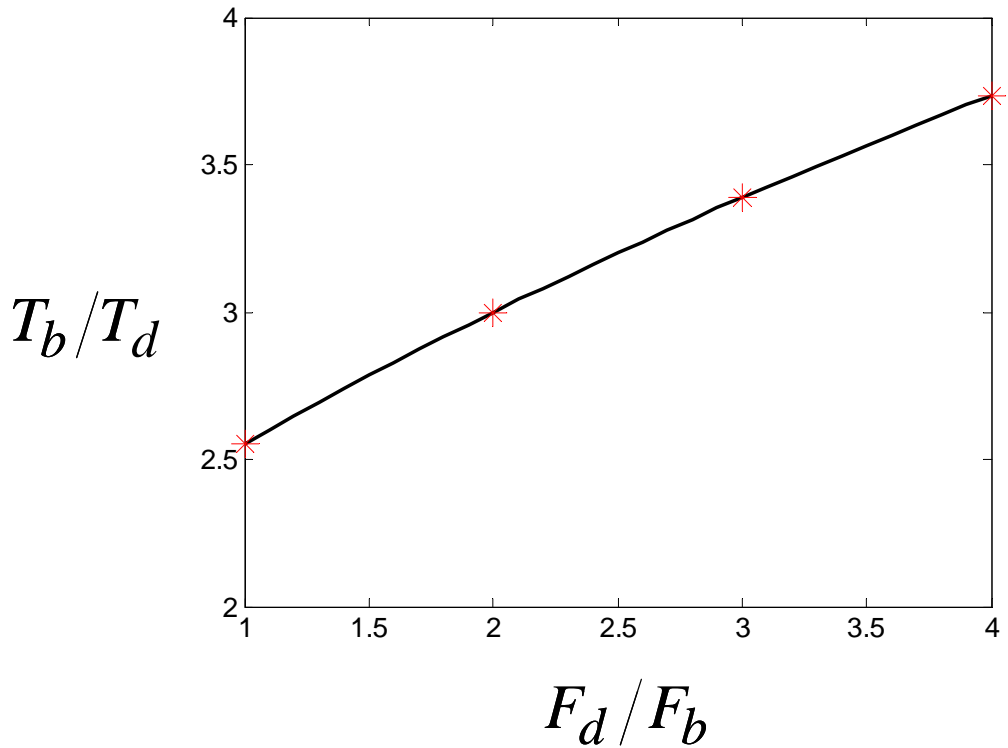

Figure 2: The ratio of backswing time to downswing time as a function of the ratio of downswing force to backswing force as derived from the simple harmonic oscillator. The ratio $T_b/T_d$ falls in the range $2.5 < T_b/T_d < 3.5$ as long as the ratio $F_d/F_b$ is constrained to $1 < F_d/F_b < 3$.

While this simple model can not possibly capture all of the dynamics of the golf swing, it has two very attractive aspects. First, the time of the backswing is not dependent on the applied force; rather, it is defined by the resonant frequency of the system. This is consistent with the notion that the duration of the backswing is independent of the length of the shot. Second, the resulting values of $T_b/T_d$ are comparable to the measured values, lending additional plausibility to this model.

We have conducted the following experiment to further test the hypothesis that at its core, the golf swing is the biomechanical equivalent of a simple harmonic oscillator.

The experiment is designed to measure the motion of the torso during the backswing swing subject to a constant applied force. In particular, the experiment tests if the angle of rotation of the torso can be used as the generalized coordinate for the oscillator. As was derived above, the simple harmonic oscillator model suggests the following two hypotheses: 1) the duration of the backswing should be independent of the applied force and 2) the amount of torso rotation should increase linearly with applied force.

In the experiment a golfer is asked to hit golf balls, shots of varying length from relatively short chip shots to the longest possible shot. While in the address position, one end of a cable is attached to the handle of the golf club, the other end of the cable fastened to a load cell (Transducer Techniques, model # MLP-50 ), and the cable is oriented along the target line so as to oppose the backswing.

As the golfer tries to initiate the golf swing, the cable assembly prevents the backswing but measures the initial force applied by the golfer at the start of the swing. Once the applied force is measured, of order one to two seconds, the cable is cut and the golf swing is initiated. An electromechanical gyroscope (Analog Devices ADXRS 300), attached to the back of the golfer and oriented along the spine, is used to measure the rotation of the torso. The signals from both sensors are digitized and then collected by a computer, yielding a complete set of data every 4.4 ms. Data are taken throughout the entire golf swing and data are then taken for many golf swings. From these measurements we extract the maximum rotation of the torso during the backswing as a function of applied force and the total time of the backswing as a function of applied force.

The data for a representative golfer with "tour tempo" are summarized in Fig. 3 and Fig 4. Shown in Fig. 3 is the time of the backswing as a function of the applied torque. The torque is calculated having determined that the force measured by the load cell is applied through a moment arm of length 0.47 m normal to the axis of rotation of the torso.

The solid line represents the mean of the measurements. It is clear that the duration of the backswing is relatively insensitive to the applied torque, consistent with our first hypothesis.

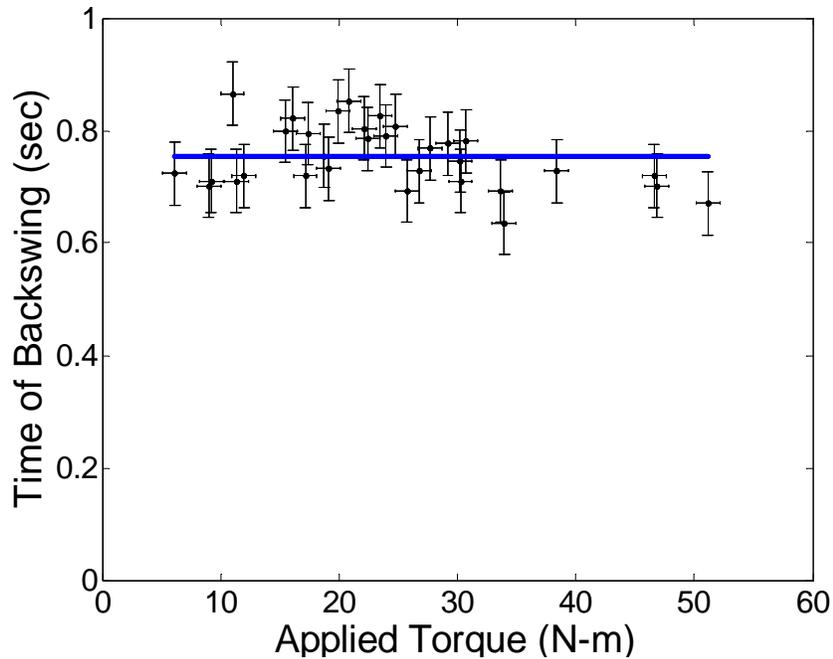

Figure 3. The time of the backswing as a function of applied torque. The solid line indicates the mean of the measurements, 0.75 sec.

Fig. 4 shows the maximum rotation of the torso as a function of the applied torque. From the data in Fig. 3, we can make an estimate of the maximum torso rotation by assuming the rotation angle of the torso is the relevant displacement parameter for our harmonic oscillator model. From the time of the backswing, we obtain an estimate of the resonant frequency, $\Omega = \frac{\pi}{T_b}$. With an estimate of the rotational inertia, one can then make a guess as to the spring constant, $k_\theta = I\Omega^2$.

Biomechanics allows us to estimate the inertia of the system, which we do for the case of the golfer in the address position. Our estimate includes the inertia of the golf club, torso, arms, and thighs. The rotational inertia of the body is calculated using the parameters of de Leva [9]. Rotation is assumed to occur around the long axis of the torso. The orientation of the arms and club relative to the torso correspond to the stance of the golfer under study. .

The mass of the golf club is $m = 0.43$ kg. The first and second moments of the club were measured relative to a point on the shaft that is between the left and right hands when the club is held by the golfer. The first moment is measured to be 0.27 kg-m and the second moment 0.22 kg-m².

The resulting rotational inertia of the system is calculated to be $I \sim 1.55$ kg-m² with an accuracy of order 10%. This yields an estimate for $k_\theta = I\left(\frac{\pi}{T_b}\right)^2 \sim 27$ N-m/radian. One can estimate the resulting maximum rotation using the expression $\theta_{max} = \frac{2\tau_b}{k_\theta}$. This expression is shown as the solid blue line in Fig. 4. Note that this estimate agrees relatively well with the data at low torque, but the maximum rotation angle clearly saturates at higher torque.

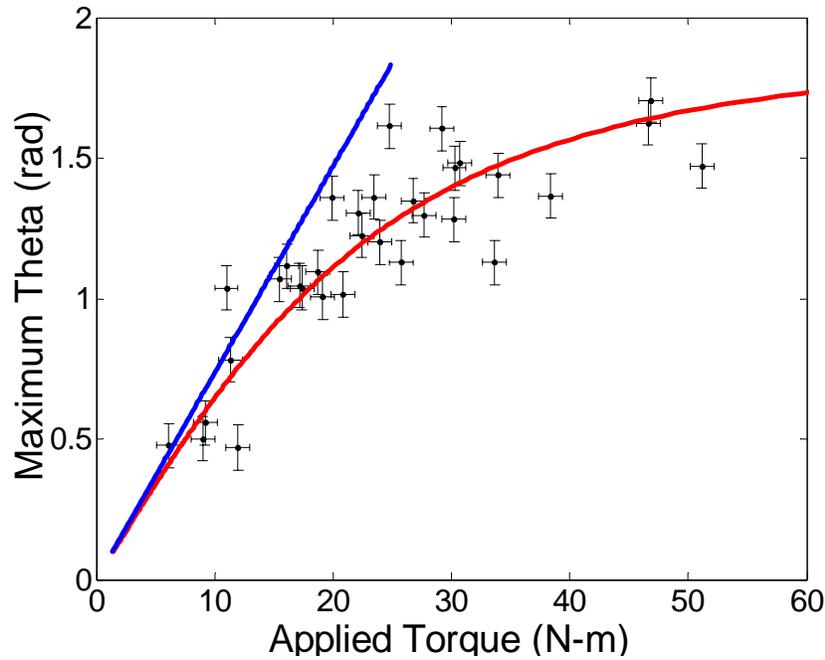

Figure 4: Maximum angle of rotation of the torso as a function of the applied torque in the backswing. The solid blue line is an estimate of what is expected if the torso responds as a simple spring. The solid red curve is a fit to a model where the torso responds like a non-linear spring.

It is well know that the torso can be modeled as a non-linear spring in which the torso becomes stiffer as the angle of rotation is increased [10, 11]. These measurements are typically done by applying a known force directly to the torso and measuring the resulting motion while the hips are restrained from movement. The resulting spring constant is generally modeled as having an exponential dependence on displacement, $k_\theta = k \exp\left(\dfrac{\theta}{\theta_0}\right)$. While this is not the exact same constraint as in the golf swing, one might imagine similar phenomena.

We have fit our data in Fig. 4 to several non-linear models. Shown in Fig. 4 as the red curve is a fit to the model $k(\theta) = \dfrac{k}{\left(1 - \dfrac{\theta^2}{\theta_0^2}\right)}$. This two parameter model, which was found to adequately fit the data, assumes the spring constant is symmetric about the origin and saturates more strongly than the exponential model described above. The primary purpose in fitting this model to the data is to extract a value of $k$, the small angle spring constant, separate from the saturation effects which are completely accounted for by $\theta_0$.

The maximum rotation angle, $\theta_{max}$, corresponds to the point at which the potential energy of the spring equals the work done by the applied torque.

$$\int_0^{\theta_{max}} k_\theta \, \theta \, d\theta = \int_0^{\theta_{max}} \tau \, d\theta.$$

Using the spring constant model above, yields the expression

$$\tau = -\frac{k\theta_0^2}{2\theta_{max}} \ln\left[1 - \left(\frac{\theta_{max}}{\theta_0}\right)^2\right],$$

where $\tau$ is the applied torque. This expression is fit to the data in Fig. 4 and is indicated by the red curve. The resulting fitting parameters are $k = 29 \pm 4$ N-m/radian and $\theta_0 = 1.8 \pm 0.3$ radians [12]. Note that this value of $k$ implies an inertia $I = 1.7 \pm 0.3$ kg-m$^2$, which is consistent with our earlier estimate of the rotational inertial at the address position. It is clear that the elastic properties of the body, whether they are passive or active, combined with the inertial properties of the body/club system provide a self-consistent, first-order explanation of the tempo of the golf swing

.The data in Fig. 4 provide clear evidence that the body functions as a non-linear spring when the torso rotation becomes large, consistent with biomechanical expectations. This non-linearity poses interesting challenges in the pursuit of a more complete analysis of the biomechanical foundation of tempo. In particular, a system with the non-linear spring described above should exhibit a decrease in the time of the backswing $T_b$ with increasing applied torque. We have modeled this phenomenon and found that $T_b$ decreased by as much as 50% for the largest applied torques, which is clearly not observed in the experiments. Implicit in this calculation is a constant applied torque and a constant rotational inertia throughout the entire backswing. In fact, neither of these conditions is likely to be true. Our experimental configuration only measures the initial applied torque and, correspondingly, our inertial calculations only consider the address position. However, there is good reason to expect the applied torque and the rotational inertia change significantly from the beginning of the swing to the top of the backswing. We propose these issues warrant a much more through series of biomechanical experiments and analysis. Indeed, given the above arguments it is intriguing that the time of the backswing is relatively insensitive to the initial applied torque and, therefore, the length of the backswing. On the face of it, one might expect these various non-linearities cancel each other so as to yield a system that, in the end, functions very nearly linearly.

In summary, the tempo of the golf swing of professional golfers exhibits remarkable uniformity in 1) the absolute time scale, 2) the ratio of backswing time to downswing time and 3) the invariance of these times as a function of the length of the swing. These observations suggest that professional golfers have at the core of their golf swing a biomechanical clock. We propose that this clock and the resulting tempo is defined by of the rotational inertia of the body/club system and the elastic properties of the body, yielding a system which can be modeled as a simple harmonic oscillator.

The results of an experiment designed to test this hypothesis are reported. The time of the backswing is measured to be independent of the applied torque while the length of the backswing increases with applied torque, consistent with this simple model. Additionally, the rotational inertia of the system (i.e. club and body) combined with the time of the backswing yield an excellent estimate of the linear component of the spring constant. Thus, to first order, we demonstrate that the tempo of the golf swing can be understood in terms of the physics of the simple harmonic oscillator.

It is observed that the rotation of the torso during the backswing is consistent with the winding up of a non-linear spring, which is consistent with biomechanical expectations. A complete biomechanical understanding of the tempo of the golf swing, including the non-linearity of the spring constant, the dynamics of the applied torque, and the position dependent rotational inertia remains an open issue and is proposed as the subject of future study.

Acknowledgements:

We would like to Peter Reeves for help in setting up some of the initial experiments; Bill Greenleaf, Mike Hebron, and David Leadbetter for valuable discussions and access to their students; and the many golfers who participated in this study.

References:

1 A. Cochran and J. Stobbs, *Search for the Perfect Swing*, (Triumph Books, Chicago, 1999).

2 J. Novosel and J. Garrity, *Tour Tempo*, (Doubleday, New York, 2004).

3 R.D. Grober, *Golf Swing Tempo Measurement System*, US Patent Application 20060063600, March 23, 2006.

4 R.M. Alexander, *Elastic Mechanisms in Animal Movement*, (Cambridge University Press, New York, 1988).

5 R.M. Alexander, *Tendon elasticity and muscle function*, Comp. Biochem. Physiol. A. Mol. Integr. Physiol. **133**(4), 1001-1011 (2002).

6 R.M. Alexander, H.C. Bennett-Clark, *Storage of elastic strain energy in muscle and other tissues*, Nature **265** (5590), 114-117 (1977).

7 R.M. Alexander, H.C. Bennett-Clark, *Storage of elastic strain energy in muscle and other tissues*, Nature **265** (5590), 114-117 (1977).

8 R.M. Alexander, *Optimum muscle design for oscillatory movements*, J. Theor. Biol. **184**, 253-359 (1997).

9 P. de Leva, *Adjustments to Zatsiorsky-Seluyanov's Segment Inertia Parameters*, J. Biomechanics **29**(9), 1223 (1996).

10 S. McGill, J. Seguin, G. Bennett., *Passive stiffness of the lumbar torso in flexion, extension, lateral bending, and axial rotation: effect of belt wearing and breath holding*, Spine **19**(6), 696 (1994).

11 A. Boden and K. Oberg, *Torque resistance of the passive tissues of the trunk at axial rotation*, Applied Ergonomics **29**(2), 111 (1998).

---

12  The uncertainties are estimated as 68% confidence limits determined by chi-square testing.  See, for example, W.H. Press, S.A. Teukolsky, W.T. Vetterling, and B.P. Flannery, *Numerical Recipes in C:  The Art of Scientific Computing*, 2nd Edition, (Cambridge University Press, New York, 1992), Chapter 15.